\def\ie{{\it i.e.},}
\def\eg{{\it e.g.},}
\documentstyle[preprint,aps]{revtex}
\begin{document}
\draft

\author{A. R. Denton} 

\address
{Institut f\"ur Festk\"orperforschung, Forschungszentrum J\"ulich GmbH, 
D-52425 J\"ulich, Germany}

\author{P. A. Egelstaff}

\address
{Department of Physics, University of Guelph, Guelph, Ontario, Canada N1G 2W1}

\title{Implicit Finite-Size Effects in Computer Simulations$^1$}

\date{\today}
\maketitle

\begin{abstract}
The influence of periodic boundary conditions (implicit finite-size effects) 
on the anisotropy of pair correlations in computer simulations is studied 
for a dense classical fluid of pair-wise interacting krypton atoms 
near the triple point. 
Molecular dynamics simulation data for the pair distribution function
$g_N({\vec r})\equiv g_N(r,\theta,\phi)$ of $N$-particle systems, as a function 
of radial distance $r$, polar angle $\theta$, and azimuthal angle $\phi$, 
are compared directly with corresponding theoretical predictions
[L.~R.~Pratt and S.~W.~Haan, J. Chem. Phys. {\bf 74}, 1864 (1981)].
For relatively small systems of $N=60$, $80$, and $108$ atoms, 
significant angular variation is observed, which is qualitatively,  
and in several cases quantitatively, well predicted by theory.
Finite-size corrections to the spherically-averaged {\it radial} 
distribution function $g_N(r)$, however, are found to be comparable to 
random statistical errors for runs of $10^5$ time steps.

\bigskip
\bigskip
\noindent
$^1$~Dedicated to Professor W. G\"otze on the occasion of
his $60^{th}$ birthday.

\end{abstract}

\bigskip
\bigskip



\section{INTRODUCTION}

Beyond the usual random statistical errors associated with averaging over 
a limited sample of particles in a computer simulation, systematic errors 
also may arise due to the finite size of the model system.  
In conventional molecular dynamics or Monte Carlo simulations of simple 
atomic systems, the size of the system usually may be chosen sufficiently 
large that finite-size effects can be safely neglected.  In contrast, 
simulations of complex molecular ({\eg} polymeric or amphiphilic) systems 
and {\it ab initio} simulations ({\eg} for electronic structure) 
must often be limited to relatively small systems, for which size 
corrections may be a practical concern.

Two general types of finite-size effect have been identified~\cite{LP1,LP2}:
(1) {\it explicit} (or ensemble) size effects, caused by the
suppression of density fluctuations upon fixing the number of particles 
(as in the canonical, microcanonical, and molecular dynamics ensembles); and 
(2) {\it implicit} (or anomalous) size effects, resulting from breaking of
orientational symmetry due to imposed (usually periodic) boundary conditions.  
Periodic boundary conditions essentially eliminate surface effects in 
simulations of bulk phases, but entail spurious correlations, which can 
measurably affect the structure of sufficiently small systems.
Both types of size effect are manifested in the detailed form of the pair 
distribution function, and thus also in any related thermodynamic properties.  
In particular, explicit size effects alter the long-range tail of the 
radial distribution function $g_N(r)$ for an $N$-particle system, whereas
implicit size effects induce anisotropy in pair correlations, as reflected 
in angular dependence of the pair distribution function.

In previous work~\cite{SDE}, we have proposed a new method to correct for 
explicit finite-size effects in extracting the static structure factor 
$S(Q)$, especially in the problematic low-$Q$ (long-wavelength) regime, 
from simulation data for simple fluids.  
That method is based on the known asymptotic 
form of the radial distribution function~\cite{LP1,LP2,Hill},  
$g_N(r)\simeq 1-S(0)/N$, and on the Fourier transform relation between 
$g_N(r)$ and $S(Q)$.
Here we focus rather on {\it implicit} finite-size effects in simulations, 
and investigate their dependence on system size.
Some 15 years ago, a detailed formal theory (and practical approximation) 
for the effects of periodic boundary conditions on equilibrium properties 
of simulated fluids was put forth by Pratt and Haan~\cite{PH1}.
At the same time, an initial, though limited, test was conducted~\cite{PH2} 
using the available molecular dynamics (MD) data of Mandell~\cite{Mandell} 
for the anisotropic structure of a dense Lennard-Jones argon fluid.
Otherwise, however, no further examination of the theory 
appears to have been documented.

This motivates the present work, the main purpose of which is to 
present an extensive numerical examination of the 
effects of periodic boundary conditions on pair correlations
in computer simulations.
To this end, we have performed a series of standard MD simulations of 
relatively small samples of pair-wise interacting liquid krypton, 
a classical simple atomic system.
Departing from standard practice, however, we have analysed the 
particle coordinates to extract the detailed {\it angular} dependence 
of the pair distribution function.  
For comparison, we have also numerically implemented the theory
of Pratt and Haan~\cite{PH1}. 
For the systems considered, significant anisotropy in pair correlations
is observed, in generally good agreement with predictions of the theory.
Although the results pertain specifically to atomic liquids, our conclusions 
are sufficiently general to be of relevance for more complex simulations. 

In the next section we first briefly outline the theory~\cite{PH1} 
of implicit finite-size effects, and then point out an implication 
for the scaling of such effects with system size.
In Sec. III, after describing the technical details of our MD simulations 
and data analysis, we present numerical results for angular variation of 
the pair distribution function and compare with theoretical predictions.  
We derive as well the implied system size dependence of the more commonly 
studied radial distribution function.
Finally, in Sec. IV we summarize and close with remarks bearing on 
the relative importance of implicit finite-size effects in 
computer simulations of more complex molecular systems.

\section{THEORY OF IMPLICIT FINITE-SIZE EFFECTS}

A formally exact theory of implicit finite-size effects in computer
simulations has been developed by Pratt and Haan (PH)~\cite{PH1}.  The theory
is based on the observation that a periodically replicated ensemble of 
simulation cells is completely equivalent to an {\it infinite} system of 
oriented ``supermolecules", each consisting of a single physical (or core) 
atom from the primary simulation cell rigidly connected to all of its 
periodic images.  Relative to the core atom, the image atoms sit 
on the sites of a lattice whose unit cell is the simulation cell.
The key point is that equilibrium statistical mechanical properties of
the finite simulated system may be described by precisely the same 
theoretical methods commonly applied to bulk molecular systems.  
In particular, the pair distribution function $g_N({\vec r})$ for the 
$N$-particle system can be related to its equivalent for the 
infinite supermolecular system.
For an atomic system with uniform one-particle density $\rho({\vec r})=\rho$, 
the pair distribution function is defined such that 
$(\rho g_N({\vec r}) d{\vec r})$ is equal to the ensemble average 
of the number of atoms in a volume element $d{\vec r}$ at position ${\vec r}$, 
in the presence of a particle fixed at the origin.  
Thus, $g_N({\vec r})$ is proportional to the conditional probability of 
finding a particle at ${\vec r}$, given a particle at the origin.
Using standard cluster expansion techniques~\cite{Stell,HM},  
$\ln g_N({\vec r})$ can be formally expressed as an exact expansion 
in terms of two-supermolecule Mayer cluster functions~\cite{PH1}.

If interactions are of sufficiently short range that
a given particle does not interact with any of its periodic images, 
as is the case when the pair potential is truncated within the 
simulation cell, then an important class of ``spring bond" graphs
can be summed to all orders~\cite{PH1}.
Working in the grand canonical ensemble to avoid explicit size effects, 
and summing exactly a certain class of ``unbridged" graphs, which depend on 
only the infinite-system function $g(r)$, but neglecting another class of
``bridged" graphs, PH have derived a practical approximate relation between
$g_N({\vec r})$ and $g(r)$, which may be expressed in the superposition-like 
form
\begin{equation}
g_N({\vec r}_{12})\simeq g(r_{12})\prod_i g(|{\vec r}_1-{\vec r}_{2i}|),
\label{PH}
\end{equation}
where
${\vec r}_{12}\equiv {\vec r}_1-{\vec r}_2$ and the product index runs over
all periodic images of particle 2.  
Physically, Eq.~(\ref{PH}) fulfills the intuitive expectation that the
statistical correlation of a given particle 1 with a second particle 2
depends not only on the position of particle 2, but also on the positions
of all periodic images of particle 2.
Evidently, $g_N({\vec r})$ is anisotropic to the extent that distances 
between image particles fall appreciably within the range of 
pair correlations in the corresponding infinite system.
Expansion (\ref{PH}) is exact to first order in the bulk density and, 
since the neglected graphs are relatively highly connected, 
the approximation is expected to be accurate for sufficiently
large system sizes and short-ranged interactions.  
Numerical computation of $g_N({\vec r})$ from Eq.~(\ref{PH})
clearly requires prior knowledge of the bulk function $g(r)$.  
Practical implementation of the theory is discussed in Sec.~III B.

The characteristic dependence of $g_N({\vec r})$ on system size $N$ 
implied by the PH theory can be inferred by the following argument.
The size dependence clearly arises because the image distances
$|{\vec r}_1-{\vec r}_{2i}|$ depend implicitly on the cell length $L$.
Suppose that $L$ is large enough that the {\it shortest} image distance
already lies in the long-range asymptotic tail of $g(r)$.  
Then, in terms of the pair correlation function $h(r)\equiv g(r)-1$, 
we can write [from Eq.~(\ref{PH})], 
\begin{eqnarray}
g_N({\vec r}_{12})/g(r_{12})&=&\prod_i\Bigg[1+h(|{\vec r}_1-{\vec r}_{2i}|)
\Bigg]\nonumber\\
&\simeq&1+\sum_i h(|{\vec r}_1-{\vec r}_{2i}|),
\label{gn1}
\end{eqnarray}
to leading order in $h(r)$. 
Therefore, the system size dependence of $g_N({\vec r}_{12})$ is dictated 
by the asymptotic behaviour of $h(r)$.
Suppose, for example, that $h(r)$ has power-law asymptotic 
behaviour~\cite{Enderby} of the form $h(r)\sim 1/r^n$ ($r \to \infty$).
Since each term in the sum then scales with system size as $1/L^n$
(to leading order), and since $L\sim N^{1/3}$, we conclude that
\begin{equation}
g_N({\vec r}_{12})/g(r_{12}) \sim 1+O\Bigg(\bigg(\frac{1}{N}\bigg)^{n/3}\Bigg).
\label{gn2}
\end{equation}
Thus, assuming power-law asymptotic pair correlations,  
the PH theory predicts that for $n>3$ implicit finite-size corrections 
to $g_N({\vec r})$ decrease faster with system size than do explicit 
$O(1/N)$ corrections.  Of course, for more rapidly ({\eg} exponentially)
decaying pair correlations the size corrections would decrease
correspondingly faster.

\section{MOLECULAR DYNAMICS SIMULATIONS}

\subsection{SIMULATIONS AND ANALYSIS}

In order to test the theory described in Sec. II, 
we have performed isothermal-isochoric MD simulations for systems of 
$N=60$, $80$, and $108$ atoms of krypton, a well-studied experimental 
system~\cite{krypton}. 
Periodic boundary conditions were applied to a cubic simulation cell
of length $L$, 
such that a particle leaving the cell through one face was simultaneously 
replaced by an image particle entering through the opposite face.
The atoms were assumed to interact via the Aziz pair potential~\cite{Aziz},  
which is known to give an accurate description of the structure and 
thermodynamics of rare-gas systems.  For krypton, the pair potential is 
parametrized by the distance at which it crosses zero, $\sigma_o=3.579 \AA$, 
the distance at which it attains a minimum, $\sigma_m=4.012 \AA$, 
and the well depth $\epsilon/k_B=200 K$.  
To avoid direct interaction between a given particle and its periodic images, 
the pair potential was truncated and shifted to zero at a 
cut-off distance $r_c=L/2$.
Particle trajectories were computed by solving the classical equations of 
motion with a fifth-order Gear's predictor-corrector algorithm.  
The thermodynamic state for all of our simulations is defined by the
reduced density $\rho^*\equiv\rho \sigma_o^3=0.8$ and the reduced average
temperature $T^*\equiv k_BT/\epsilon=0.7$, placing the system near its 
triple point.
Aside from an initial equilibration phase, during which velocities were 
rescaled to establish the desired average temperature, the total number 
of particles, volume, energy, and momentum were kept fixed.
Runs of $10^5$ time steps, with a time step of $0.005$ ps, generated 
about $1000$ independent configurations, sufficient to ensure that random 
statistical errors are small in comparison with the finite-size effects
of interest.

As a preliminary test of the simulation and analysis methods, we first
compute the radial distribution function $g_N(r)$, defined such that 
the quantity $\rho g_N(r)(4\pi r^2\Delta r)$ is equal to the ensemble average
of the number of atoms in a spherical shell of thickness $\Delta r$ at 
radial distance $r$ away from a particle centred at the origin ($r=0$).  
Note that the isotropic function $g_N(r)$ is the spherical average of
the anisotropic function $g_N({\vec r})$ defined in Sec. II.
Following standard practice, $g_N(r)$ is computed by binning particles 
in radial shells centred on a given central particle and then averaging 
over all central particles and over all configurations.

Figure~1 shows the resulting $g_N(r)$ for the system size $N=108$
out to a distance of $r=L/2$.
Choosing the shell thickness $\Delta r=0.028 \sigma_o$, the
random statistical errors are at most $\pm 0.02$.  
Corresponding results for $N=60$ and $N=80$ are found to be 
practically identical to within statistical error, suggesting that 
the function $g_N(r)$ is a reasonable measure of the bulk $g(r)$, 
and that at this level finite-size effects are relatively minor.  
We return to this point below.
Also shown in Fig.~1, for comparison, is the prediction of the perturbation 
theory of Weeks, Chandler, and Andersen (WCA)~\cite{WCA}, according to which
\begin{equation}
g(r)\simeq \exp[-u_o(r)/k_BT] y_{PY}(r/d;\rho d^3).
\label{WCA}
\end{equation}
Here $u_o(r)$ is the repulsive part of the pair potential, $y_{PY}$
is the Percus-Yevick (PY) indirect correlation (or cavity) function of the
hard-sphere fluid, and $d$ is the WCA prescription~\cite{WCA} for the 
effective hard-sphere diameter.  At the thermodynamic state investigated, 
$d=1.027 \sigma_o$.
For $y_{PY}$, we use the analytic solution of the PY equation for 
hard spheres~\cite{HM,PAE}.
The WCA prediction is seen to be in reasonable agreement with the 
simulation data, despite minor discrepancies around the first and 
second peaks.

As discussed in Secs.~I and II, pair correlations in a finite system are 
{\it not} necessarily isotropic [{\ie} $g_N({\vec r})\neq g_N(r)$] 
because of broken orientational symmetry.  
As a measure of anisotropy, we have examined the angle-dependent 
pair distribution functions
\begin{equation}
g_N(r,\phi)\equiv {1\over{2}}\int_{-1}^1 d(\cos\theta) g_N({\vec r})
\label{gphi}
\end{equation}
and 
\begin{equation}
g_N(r,\theta)\equiv {1\over{2\pi}}\int_0^{2\pi} d\phi g_N({\vec r}),
\label{gth}
\end{equation}
where $g_N({\vec r})\equiv g_N(r,\theta,\phi)$, with $r$, $\theta$, and 
$\phi$ being the radial distance, polar angle, and azimuthal angle.  
Note that the radial distribution function is obtained by integrating 
either Eq.~(\ref{gphi}) with respect to $\phi$ or Eq.~(\ref{gth}) 
with respect to $\cos\theta$.
By convention, the Cartesian axes specified by ($\theta=\pi/2, \phi=0$), 
($\theta=\pi/2, \phi=\pi/2$), and ($\theta=0, \phi$ arbitrary) are
directed perpendicular to the sides of the simulation cell.
Symmetry dictates that $g_N(r,\phi)$ is periodic in $\phi$ with period
$\pi/2$ and is mirror symmetric about $\phi=\pi/4$.
Similarly, $g_N(r,\theta$) is periodic in $\theta$ with period $\pi$
and is mirror symmetric about $\theta=\pi/2$.
All angles therefore can be represented in the reduced ranges 
$0\leq\phi\leq\pi/4$ and $0\leq\theta\leq\pi/2$.
In practice, $g_N(r,\phi)$ is computed by counting the number of particles 
within a {\it particle-centered} longitudinal wedge
of longitude $\phi$, radius $r$, angular width $\Delta\phi$, 
and radial thickness $\Delta r$, and then averaging over all 
central particles and all configurations.
Similarly, $g_N(r,\theta$) is computed by counting the number of particles 
within a latitudinal slice of latitude $\theta$, 
radius $r\sin\theta$, angular width $\Delta\theta$, and radial thickness
$\Delta r$, averaged over all central particles and configurations.

\subsection{RESULTS AND COMPARISON WITH THEORY}

Figures~2-4 display our MD results for the angle-dependent distribution 
function $g_N(r,\phi)$, as a function of azimuthal angle $\phi$ 
[Eq.~(\ref{gphi})],  
at fixed radial distances $r/\sigma_o=1.1, 1.6$, and $2.1$ -- near the 
first maximum, the first minimum, and the second maximum of $g_N(r)$, 
respectively (see Fig.~1).
For the smallest system (Fig.~4), only the data for the first two distances 
are shown, since the cut-off at $r=L/2$ occurs before the second maximum 
of $g_N(r)$.
Figure~5 presents similar results (largest system only) for 
$g_N(r,\theta)$ as a function of polar angle $\theta$ [Eq.~(\ref{gth})]. 
Note that as $\theta \to 0$ the statistical error increases because of 
the geometrical factor $\sin\theta$ in the volume of the latitudinal slice.
In order to accumulate reasonable statistics, we have used angular widths
$\Delta\phi=\Delta\theta=\pi/36$ and a radial thickness 
$\Delta r=0.112 \sigma_o$.
As the latter is four times larger than the shell thickness used in 
the calculation of $g_N(r)$, the average values of $g_N(r,\phi)$ and 
$g_N(r,\theta)$ differ slightly from the corresponding values of $g_N(r)$.
The angular variations illustrated in Figs.~2-5 are clearly significant --
up to five times larger than the statistical error in the case of $\phi$, 
and up to ten times larger in the case of $\theta$.
We mention in passing that we have also analysed the one-particle density 
$\rho({\vec r})$ and confirmed it to be translationally uniform, {\ie}
$\rho({\vec r})=\rho$. 

Also shown in Figs.~2-5, for comparison, are corresponding predictions of 
the PH theory.  As noted in Sec.~II, numerical implementation of the theory 
requires knowledge of the bulk system radial distribution function $g(r)$.
To allow for a consistent comparison between theory and simulation, 
for distances $r<L/2\simeq 2.56 \sigma_o$ we approximate the required 
infinite-system $g(r)$ by its finite-system counterpart $g_N(r)$ 
computed from our $N=108$ simulation. 
This represents a good approximation to the true bulk function, 
as shown in Sec. II and discussed further below.
For larger distances $r>L/2$, beyond the range of the simulation data, 
we use the $g(r)$ predicted by the WCA perturbation theory~\cite{WCA}. 
As demonstrated in Sec. II, this gives a reasonable fit for the
relatively dense system considered here (see Fig.~1). 
(We note, however, that at lower densities, where perturbation theories 
tend not to perform as well, more accurate integral-equation theories 
may be required.)
The function $g_N({\vec r})$ is computed directly from Eq.~(\ref{PH}), 
taking into account all image particles in a 5x5x5 ensemble of cells
centred on the primary cell ($124$ images).
Numerical integration with respect to $\theta$ or $\phi$ then yields
the distribution functions defined by Eqs.~(\ref{gphi}) and (\ref{gth}).
Evidently, the PH theory tracks the angular variations remarkably well, 
at least at the thermodynamic state that we have simulated. 
Where quantitative discrepancies occur, the theory tends to 
underestimate the magnitude of the variations.
As a check on the calculations, note that $g_N(r,\phi=0)=g_N(r,\theta=\pi/2)$, 
as should be so by symmetry.

Finally, we examine the effect of periodic boundary conditions on 
the radial distribution function by computing the PH prediction 
for the spherically-averaged function
\begin{equation}
\delta g_N(r) = \bigg[{1\over{4\pi}}\int_0^{2\pi}d\phi\int_{-1}^1 d(\cos\theta)
g_N({\vec r})\bigg]-g(r).
\label{dgnr}
\end{equation}
This gives a measure of the deviation of the $N$-particle function $g_N(r)$ 
from its bulk counterpart $g(r)$.  
For simplicity, we use in this case the WCA $g(r)$ for all distances.
As Fig.~6a illustrates, the predicted $\delta g_N(r)$ oscillates as a 
function of $r$ about zero with an amplitude of less than $0.02$ for $N=108$, 
increasing as the system size decreases.
The oscillations arise from the fact that as $r$ increases the image distances
$|{\vec r}_1-{\vec r}_{2i}|$ in Eq.~(\ref{PH}) alternate between regions where
$g(r)$ is greater or less than unity.
The amplitude is in fact comparable to the random statistical errors 
in our simulations (see Fig.~1), although of course for longer runs it would 
begin to exceed statistical error.  
Interestingly, at distances near the first maximum, first minimum, and
second maximum ($r/\sigma_o=1.1$, $1.6$, and $2.1$) the deviation 
from bulk behaviour essentially vanishes.
This further justifies our use above of the MD $g_N(r)$ as an approximation 
to the bulk $g(r)$ in numerically implementing the PH theory
to study angular variations.
It should be emphasized that the practical utility of the PH theory for 
predicting finite-size effects on thermodynamic properties related to 
volume integrals of $g(r)$, such as the equation of state, 
is limited by an inherent thermodynamic inconsistency in the 
approximation of Eq.~(\ref{PH})~\cite{Kolafa}.
In this connection, however, exact implicit finite-size corrections 
to virial coefficients have been derived~\cite{Kratky}.

The predicted scaling of $g_N(r)$ with system size is shown in Fig.~6b 
at two fixed distances.  
The oscillations arise from the alternation of image distances,  
with increasing $L$, between regions where $g(r)>1$ and regions 
where $g(r)<1$, and have a wavelength increasing as $N^{1/3}$.
Clearly the $N$-particle function decays 
rapidly to its bulk limit with increasing $N$.  The functional form 
is revealed by the log-log plot (inset) of the maxima and minima 
of the bounding envelope, indicating a near-linear relation. 
For the short-ranged pair potential used here, this implies a simple 
power law of the form $g_N(r)/g(r)\sim 1+O(1/N^{\nu})$ (with $\nu\simeq 2.2$), 
consistent with the scaling argument presented in Sec.~II.

\section{SUMMARY AND CONCLUSIONS}

In summary, we have studied the magnitude of finite-size effects 
induced by periodic boundary conditions through a series of 
molecular dynamics simulations of a dense krypton fluid of 
between $N=60$ and $N=108$ atoms near its triple point.  
As a measure of anisotropy, 
we have analysed the variation with azimuthal and polar angles of 
the pair distribution function at several radial distances and system sizes.
Significant anisotropy was observed, the magnitude of angular variations 
being several times larger than random statistical errors
for runs of $10^5$ time steps.

For comparison, the theory of Pratt and Haan~\cite{PH1} was implemented 
and was found to satisfactorily model the anisotropies observed in 
the simulations.
The same theory predicts that spherical averaging of the 
pair distribution function considerably diminishes the magnitude of 
implicit size corrections, reducing them to the level of our 
statistical errors. 
The theory further implies that for asymptotic pair correlations 
of the power-law form $(\sim 1/r^n)$ the size correction scales with
system size as $O(1/N^{n/3})$.
This is confirmed by our comparison of simulation data for the 
radial distribution function at different system sizes, which 
indicates that implicit corrections decrease rapidly with system size, 
significantly more rapidly than explicit corrections.

We conclude by pointing out that since implicit finite-size effects can
obviously cause significant anisotropy in pair correlations even in the
case of a simple atomic liquid with spherically symmetric pair potential, 
they are certainly also a potential source of error in extracting 
orientational correlation parameters from simulations of molecular systems, 
especially for anisometric molecules~\cite{Impey}.
In MD simulations of 256 CS$_2$ molecules~\cite{Tildesley},  
for instance, significant anisotropy in the Kirkwood $g_2$ parameter
was observed~\cite{Impey}. 
Furthermore, the distortion of three-particle and higher-order correlations 
by boundary conditions is expected to be even more pronounced~\cite{PH2}.
Finally, we emphasize that here we have considered only the case of
short-range interactions.
The case of long-range interactions is relevant to simulations of
molten salts, charge-stabilized colloids, and other charged 
systems~\cite{Adams}.
Future study of these issues, by simulation and by practical extension 
of the theory of finite-size effects would be worthwhile.

\acknowledgements
\noindent
One of us (ARD) thanks N. W. Ashcroft for an inspiring discussion.
The Forschungszentrum J\"ulich is gratefully acknowledged 
for the use of computer facilities.
This work was supported in part by a grant from the Natural Sciences 
and Engineering Research Council of Canada.






\newpage
\unitlength1mm
\begin{figure}
\noindent
\caption[]{
MD simulation data (dots) and WCA prediction~\cite{WCA} (curve) 
for the radial distribution function of $108$ krypton atoms at reduced density 
$\rho^*=0.8$ and reduced temperature $T^*=0.7$.  
Error bars in the inset represent random statistical errors 
of plus or minus one standard deviation.
}
\label{FIG1}
\end{figure}

\begin{figure}
\noindent
\caption[]{
MD data (symbols) compared with theoretical predictions~\cite{PH1}
(curve) for the angle-dependent pair distribution function 
$g_N(r,\phi)$ vs. azimuthal angle $\phi$ (in radians, with $0<\phi<\pi/4$) 
for $N=108$ and radial distances (a) $r/\sigma_o=1.1$, (b) $1.6$, and 
(c) $2.1$.
Error bars represent random statistical errors 
of plus or minus one standard deviation.
}
\label{FIG2}
\end{figure}

\begin{figure}
\noindent
\caption[]{
Same as Fig.~2, but for $N=80$.
}
\label{FIG3}
\end{figure}

\begin{figure}
\noindent
\caption[]{
Same as Fig.~2, but for $N=60$.
}
\label{FIG4}
\end{figure}

\begin{figure}
\noindent
\caption[]{
MD data (symbols) compared with theoretical predictions~\cite{PH1}
(curve) for the angle-dependent pair distribution function 
$g_N(r,\theta)$ vs. polar angle $\theta$ (in radians, with $0<\theta<\pi/2$) 
for $N=108$ and radial distances (a) $r/\sigma_o=1.1$, (b) $1.6$, and 
(c) $2.1$.
}
\label{FIG5}
\end{figure}

\begin{figure}
\noindent
\caption[]{
Theoretical predictions~\cite{PH1} for the $N$-particle 
radial distribution function $g_N(r)$: (a) deviation from the bulk $g(r)$
vs. radial distance $r<L/2$ for $N=108$ (solid curve), $N=80$ (long-dashed), 
and $N=60$ (short-dashed); (b) scaling with system size $N$ for 
$r=1.6 \sigma_o$ (solid) and $r=2.1 \sigma_o$ (dashed); inset shows
scaling of bounding envelope on a log-log scale for $r=1.6 \sigma_o$ (circles) 
and $r=2.1 \sigma_o$ (squares).
}
\label{FIG6}
\end{figure}


%
%
%

\end{document}